# Motivational Climate Effects on Communications, Emotional-Social States, and Performance in Collaborative Gaming Environment

Omer Eldadi,[1] Yarin Dekimhi,[2] and Gershon Tenenbaum[1,3]

1. B. Ivcher School of Psychology, Reichman University, Herzliya, Israel
2. The College of Management, Rishon Lezion, Israel
3. Department of Physiotherapy, Faculty of Health, Ariel University, Shomron, Israel

**Corresponding Address**:

Omer Eldadi

B.Ivcher School of Psychology

Reichmann University

The University 8, Herzliya, Israel

Email: Omereldadi@gmail.com

**Abstract**

The study explores the effects of motivational climate on communication features, emotional states, collective efficacy, and performance in collaborative gaming environments. Forty participants with no prior gaming experience were randomly assigned to 20 gender-matched teams of three (including one confederate) across two motivational climates: positive-supportive (PS) or neutral-unsupported (NU) (10 teams per condition). Team members completed three progressively difficult levels of Overcooked! 2 during which communication contents, emotional responses, collective efficacy, and performance outcomes were observed and coded. Mixed-design MANOVAs and ANOVAs were employed to examine the effects of motivational climate and task difficulty on communication patterns, emotions, collective efficacy, and performance. Chi-square analyses were performed to test communication content differences between conditions. Results revealed that PS team members significantly outperformed NU teams at lower task difficulty level, but this advantage diminished as task complexity increased. Communication analysis revealed that PS team members utilized significantly more action-oriented, factual, and emotional/motivational statements, while NU team members used more statements of uncertainty and non-task-related communication. The percentage of the talk time increased with difficulty across both climate conditions. PS team members maintained more positive emotional profiles throughout, with higher excitement and happiness scores and lower anxiety, dejection, and anger compared to NU team members. Furthermore, PS team members reported consistently higher collective efficacy beliefs across all difficulty levels. These findings reveal that positive motivational climate enhances team communication effectiveness, emotional resilience, and performance outcomes in challenging collaborative environments.

**Motivational Climate Effects on Communications, Emotional-Social States, and Performance in Collaborative Gaming Environment**

Team performance in collaborative environments depends significantly on the communicative dynamics and emotional states shared among team members (Eldadi & Tenenbaum, 2025; Eldadi et al., 2023; Freeman & Wohn, 2017; Marlow et al., 2018). Effective teamwork transcends individual expertise, requiring coordinated action and information exchange under pressure (Eccles & Tenenbaum, 2004; Filho & Tenenbaum, 2020). The relationship between team communication, emotional states, and performance becomes particularly salient in contexts with increasing complexity and time constraints. The present study explores the relationships between motivational climate and team dynamics in collaborative gaming environments, examining how variations in supportive versus neutral team environments may be associated with communication patterns, emotional responses, collective efficacy, and performance outcomes.

## Theoretical Background

### Motivational Climate in Team Settings

Motivational climate represents the psychological environment created by significant others that shapes individuals' goal orientations and behavioral patterns (Ntoumanis & Biddle, 1999). Research findings identified two primary motivational climate dimensions: mastery (task-involving) climate, which emphasizes personal improvement and effort, and performance (ego-involving) climate, which prioritizes outperforming others and demonstrating superior ability (Duda & Balaguer, 2007; Nicholls, 1984).

Motivational climate theory has evolved substantially since Nicholls' (1984) seminal work on achievement goal orientations, which proposed that individuals' motivation in achievement contexts stems from different conceptions of competence and success. Building on this foundation, Ames (1992a,b) developed the TARGET framework (Task, Authority,

Recognition, Grouping, Evaluation, and Time), which identifies specific environmental structures that shape motivational climate perceptions. This framework has been extensively validated in sport and physical activity contexts, with research revealing that motivational climate consists of six first-order factors that aggregate into two second-order dimensions: task-involving (mastery) climate and ego-involving (performance) climate (Newton et al., 2000). These distinct climates create fundamentally different psychological environments that influence participants' cognitive, affective, and behavioral responses.

Mastery-oriented climates emphasize personal improvement, effort recognition, skill development, and viewing mistakes as learning opportunities, while performance-oriented climates prioritize normative comparisons, winning, and demonstrating superior ability relative to others (Duda & Balaguer, 2007). Extensive research has documented differential outcomes associated with these climates. In their comprehensive systematic review, Harwood et al. (2015) found that mastery climate perceptions were consistently associated with adaptive motivational outcomes including enhanced perceived competence, self-esteem, objective performance, intrinsic motivation, positive affective states, and ethical attitudes. Conversely, performance climates have been linked to maladaptive outcomes such as increased anxiety, fear of failure, and negative affect, particularly when individuals face challenging tasks (Ntoumanis & Biddle, 1999). Field-based intervention studies provide causal evidence for these associations, with Theeboom et al. (1995) demonstrating that children randomly assigned to mastery-oriented programs reported significantly higher enjoyment and exhibited superior motor skill development compared to those in performance-oriented programs.

**Contextual Adaptation of Motivational Climate Theory**

While this traditional dichotomy has proven valuable across numerous sport and physical activity contexts, researchers increasingly recognize that motivational climate manifestations may vary across different achievement settings. Harwood et al. (2015) emphasized that climate perceptions must be appropriately contextualized to specific achievement environments, noting that motivational climate has been studied in several physical activity contexts requiring tailored conceptualizations that capture essential psychological mechanisms while remaining sensitive to contextual demands. This recognition opens the door for thoughtful adaptations of motivational climate theory when studying novel environments, provided such adaptations maintain theoretical coherence with the underlying achievement goal framework while addressing unique contextual features.

While traditional motivational climate research examines comprehensive environmental structures across the TARGET dimensions, the unique constraints of collaborative gaming environments necessitate a more focused approach. In Overcooked! 2, where teams work cooperatively without direct competition against other teams, the traditional performance climate emphasis on normative comparison and winning becomes less relevant. Instead, the critical environmental factor that can vary is the degree of supportive communication and encouragement team members experience during collaborative task execution.

We therefore examined how different communication patterns, specifically positive-supportive (PS) versus neutral-unsupported (NU), function as key elements within the broader motivational climate of collaborative gaming. Rather than manipulating the full motivational climate, our study focuses on how variations in peer communication support, which represents one crucial aspect of the recognition and evaluation dimensions of the TARGET framework, influence team dynamics in environments where task mastery rather than competitive superiority defines success.

This approach aligns with recent perspectives suggesting that motivational climate effects may manifest differently across contexts. In collaborative gaming, where structural elements like task design and grouping are predetermined by the game itself, the interpersonal communication dimension becomes a particularly salient avenue (Eldadi & Tenenbaum, 2025) through which motivational support versus neutrality can be expressed. Our positive support (PS) condition thus incorporates key supportive elements associated with mastery climates—emphasis on effort recognition, learning from mistakes, and collective improvement—while our Neutral-unsupportive (NU) condition removes these supportive elements without introducing competitive or ego-involving components that would be incongruent with the collaborative task structure.

**Positive-Supportive versus Neutral-Unsupported Climates**

In the current study's collaborative gaming environment, teams strived to improve their own performance across increasing difficulty levels without direct competition against other teams, making the traditional ego-oriented climate less relevant. The PS climate condition better captures the essence of mastery climate in this self-referenced performance context, where supportive communication fosters team resilience and collaborative problem-solving. Avcı et al. (2018) found that athletes who perceived a mastery-oriented motivational climate reported higher levels of closeness, commitment, and complementarity in their communication with coaches, suggesting that supportive environments focused on improvement rather than competition foster better team dynamics. The PS/NU distinction retains the fundamental psychological mechanisms of achievement goal theory while being appropriately adapted to the current experimental context where success is defined in terms of task mastery rather than normative comparison, the focus is on team cooperation rather than between-team competition, and the challenges are progressively difficult tasks rather than competitive opponents.

The PS condition in our study incorporates key elements that Harwood et al. (2015) identified as central to mastery climates, including emphasis on effort, learning, skill improvement, cooperative learning, and viewing mistakes as part of the learning process. Conversely, the NU condition, while not fully representing a traditional ego climate, provides an appropriate contrast by withholding the supportive elements that characterize the PS condition, thus allowing us to examine the specific contributions of positive and supportive climate elements to team functioning in the collaborative gaming paradigm. PS and NU climates will affect differently the team members' verbal communications, and consequently their emotions, collective efficacy and their collaborative performance.

**Communication Dynamics in Team Performance**

Effective communication enhances the team's coordination performance and strengthens collective efficacy (Marlow et al., 2018). Open information exchange and collaborative idea generation facilitate the development of shared mental models and result in a more efficient coordination mechanisms (Freeman & Wohn, 2017; Mesmer-Magnus et al., 2021; Musick et al., 2021). However, the specific effect of motivational climate on communication content characteristics and their subsequent impact on team performance remains understudied.

Lausic et al.'s (2009) has identified several categories of verbal communication content that reflect both shared tactical understanding and complementary role-specific knowledge. For instance, action statements typically convey shared strategic plans, while factual statements communicate specialized knowledge about environmental conditions. In electronic games setting, expert gaming players employ more action-oriented and factual communication than novice teams members (Eldadi & Tenenbaum, 2025), providing evidence that adaptive communication patterns significantly influence performance outcomes.

In collaborative gaming contexts, where rapid decision-making, adaptability, and player coordination are essential, communication dynamics serve as a critical determinant of team performance (Musick et al., 2021). Team members employing effective communication strategies, characterized by information sharing, collaborative problem-solving, and timely feedback, consistently outperform those with less effective communication (Eldadi & Tenenbaum, 2025; Leavitt et al., 2016). This effect becomes particularly pronounced in complex, fast-paced gaming environments that require continuous adaptation.

Effective communication also functions as the primary mechanism through which team members develop and maintain team shared mental models (TSMMs; DeChurch & Mesmer-Magnus, 2010; Eccles & Tenenbaum, 2004; Eldadi & Tenenbaum, 2025; Lines et al., 2022). TSMMs represent the "extent and accuracy of shared and complementary knowledge types held by team members about the individuals in the team, team tasks, the team as a whole, and contextual constraints" (Filho & Tenenbaum, 2020, p.5). Effective team members operations consist of both shared mental models (e.g., overlapping knowledge) and complementary mental models (e.g., distributed, specialized knowledge) that enable coordinated action (Filho & Tenenbaum, 2020; Klimoski & Mohammed, 1994; Mathieu et al., 2000) and subsequently enhance performance outcomes.

**Motivational Climate Effects on Emotions**

Motivational climate profoundly shapes team members' emotional experiences. Mastery-oriented climates were found to be associated with reduced anxiety, enhanced enjoyment, and increased intrinsic motivation (Smith et al., 2007). In contrast, performance-oriented climates tended to generate greater performance anxiety, fear of failure, and negative affect, particularly when team members were confronted with challenging tasks (Ntoumanis & Biddle, 1999).

Peña and Hancock (2006) analyzed socioemotional and task communication in online gaming teams and found that team members communication patterns directly influence the emotional climate of the team, which subsequently affects intra-team coordination and performance.

The bidirectional relationship between communication and emotions is particularly evident in high-pressure scenarios where positive emotional states facilitate coordination while effective communication helps maintain those positive states (Eldadi et al., 2023).

Emotional contagion has also been linked to team resilience and adaptability in the face of challenges. Specifically, positive communication and emotional support among teammates helped foster team resilience during difficult moments in team performance (Eldadi et al., 2023; Leis et al., 2022). Moreover, positive emotional states facilitate coordination and problem-solving while enhancing team members' willingness to support one another. In high-pressure scenarios, positive emotional states among members are crucial for their performance outcomes (Eldadi et al., 2023).

**Motivational Climate Effects on Collective Efficacy and Performance**

Collective efficacy (Bandura, 1997) represents a robust social-cognitive mechanism that influences team performance. Team members with higher collective efficacy maintain greater persistence when confronting challenges and achieve superior outcomes (Goddard et al., 2004). In gaming contexts, collective efficacy functions as a psychological resource that enables team members to maintain confidence and coordination even as task difficulty increases.

The relationship between motivational climate and collective efficacy is particularly essential for enhancing team dynamics. PS climates foster stronger perceptions of collective efficacy by emphasizing controllable aspects of performance such as effort and improvement (Harwood et al., 2015). Team members in such environments develop shared confidence in their ability to overcome obstacles through persistent effort rather than through inherent ability alone.

Enhanced collective efficacy becomes especially important as task difficulty increases, providing teams with psychological resources to maintain coordination and problem-solving capabilities under pressure. Within the specific context of collaborative gaming environments like Overcooked! 2, where members must rapidly adapt to increasingly difficult challenges while maintaining coordination, collective efficacy affects team resilience and adaptive capacity (Grossman et al., 2015; Leis et al., 2022).

**Task Difficulty as a Moderator of Motivational Climate Effects**

Communication among team members changes as a function of task complexity (Marlow et al., 2018). As tasks become more complex, team communication patterns shift to accommodate increased coordination requirements. Marlow and colleagues (2018) found that task complexity moderates the relationship between communication and performance, with complex tasks demanding more structured, frequent, and explicit communication to maintain team effectiveness.

In the specific context of Overcooked! 2, which features progressively challenging levels, communication adaptations become crucial for team performance. As kitchen layouts become more complex and time constraints tighten, team members must adjust their communication strategies accordingly. Rosero et al.'s (2021) findings showed that team members playing Overcooked! 2 developed increasingly sophisticated coordination mechanisms across difficulty levels, with communication serving as the primary vehicle for this adaptation.

**A Conceptual Framework for Collaborative Gaming Environments**

Consisting of the literature reviewed, we propose a conceptual framework in which motivational climate serves as the primary independent variable simultaneously influencing multiple team processes and outcomes. Specifically, we hypothesize that motivational climate (positive-supportive - PS vs. neutral-unsupported - NU) directly influences team communication patterns, which associates with team members' emotional states, collective efficacy beliefs, and performance outcomes. Task difficulty serves as a moderator of these relationships, with the effects of motivational climate becoming more pronounced as task demands increase.

Overcooked! 2 provides an ideal platform for examining these relationships due to its cooperative design that requires players to prepare and serve meals under strict time constraints. The game's progressively challenging levels create optimal conditions for examining how teams adapt their processes in response to increasing task demands. The environment effectively simulates the coordination challenges present in many real-world team settings, where members must adjust their interaction patterns as tasks become more complex.

The study's central hypothesis is that motivational climate, specifically positive-supportive (PS) more than neutral-unsupported (NU) climate, significantly influences social-emotional-cognitive dynamics in collaborative gaming environments. We posit that these effects become increasingly pronounced as task difficulty level increases. The PS motivational climate is expected to enhance team functioning through improved communication strategies, positive emotional states, and stronger collective efficacy beliefs, ultimately yielding superior performance compared to teams in NU environments. We expect that PS team members will employ more action-oriented and factual communications across all difficulty levels than their NU counterparts and will exhibit and maintain more positive emotional states (excitement, happiness) while experiencing reduced negative emotions (anxiety, anger, dejection) compared to those in the NU condition. We anticipate this emotional-social-cognitive advantage will become particularly evident as tasks grow more challenging.

# Method

## Players

Players were recruited via university email lists and flyers posted on campus. To be eligible for the study, players were between the ages of 18 – 50 years and had no prior experience playing Overcooked! 2 or any other video game on a regular basis (0 hours per week). Sample size was determined using G*Power 3 (Faul et al., 2007), with parameters set to detect medium-to-large effects ($f = 0.35$) with 80% power at $\alpha = .05$ for the primary analyses. A total of 40 players (20 males, 20 females) were recruited and randomly assigned to 20 teams of three, each including two players and one confederate. The teams were evenly distributed across two motivational climate experimental conditions: positive-supportive (PS), and neutral-unsupported (NU), resulting in 10 teams per condition. To ensure gender balance, half of the teams in each condition included two female participants with a female confederate, while the other half consisted of two male participants with a male confederate. Teams were gender-matched to control for potential gender-based communication differences and avoid confounding effects from mixed-gender dynamics that might influence team interaction patterns independently of our motivational climate manipulation.

## The Task: Overcooked! 2

Overcooked! 2 is a cooperative cooking simulation game developed by Ghost Town Games and Team17 studios. Players work together in chaotic kitchens to prepare and serve meals within a time limit. The game features increasingly complex levels with various obstacles and kitchen layouts. Players must coordinate tasks such as chopping ingredients, cooking dishes, washing plates, and serving orders (see Figure 1 for in-game screenshot). The game's difficulty increases through more complex recipes, kitchen hazards, and tighter time constraints.

**Figure 1**

*A gameplay scene from Overcooked! 2, Task Difficulty Level Moderate (1-6')*

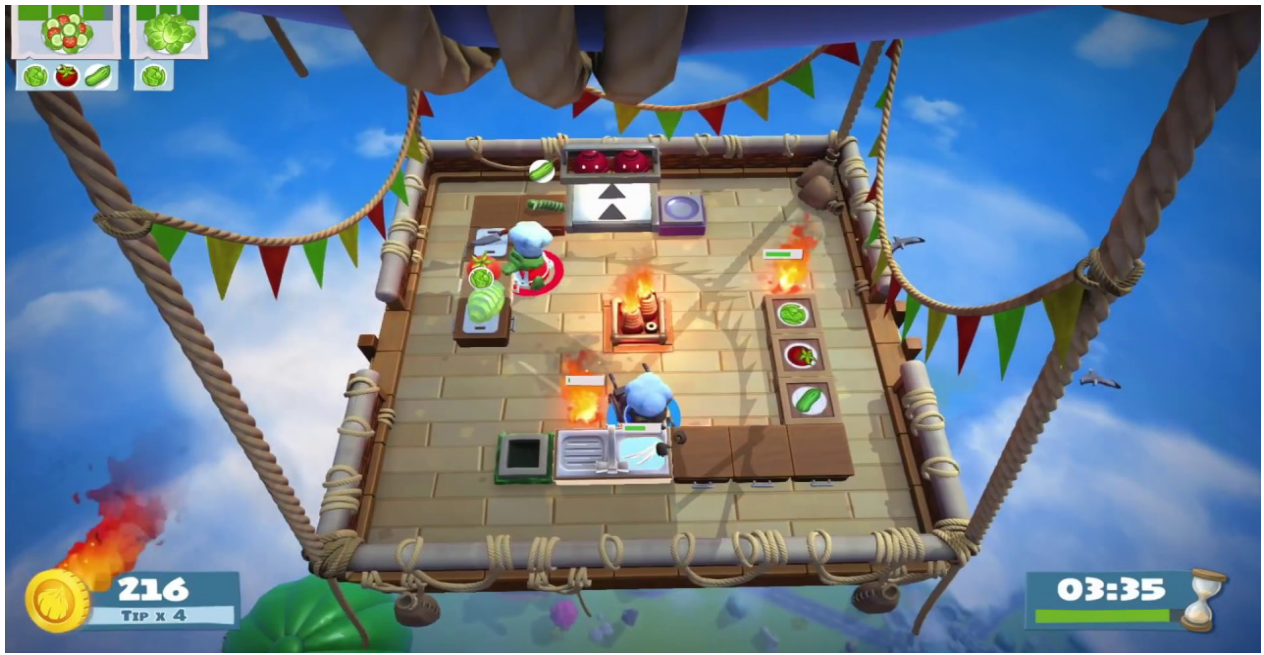

**Measurements**

      **Demographic Questionnaire** (DQ). The DQ includes items on age, gender, education level, and prior and current experience with video games.

      **Emotional Contagion** (EC; Doherty, 1997). The EC measures the psychological disposition of emotional contagion using 15 items representing five dimensions: *happiness, love, fear, sadness, and anger*. Each dimension consists of 3 items rated on a 4-point Likert-type scale, ranging from 1 (*never*) to 4 (*always*). The average rating represents the emotional intensity of each dimension. The authors reported an internal consistency of Cronbach's $\alpha$ = 0.90 and the

factor loadings ranged from 0.46 - 0.69. Positively related emotions were related to reactivity, emotionality, sensitivity to others, social function, and self-esteem. Negatively related emotions were correlated with alienation, self-assertiveness, and emotional stability. The EC was low to moderately correlated (0.30 - 0.47) with measures of responsiveness and self-reports of emotional experiences following exposure to emotional expressions. Temporal stability after a 3- week interval ranged from 0.80 to 0.82 for positive and negative emotions, respectively, and .84 for the entire scale. Women were found to be more susceptible to emotions than men.

**Sport Emotion Questionnaire** (SEQ; Jones et al., 2005). The SEQ consists of 22 items representing 5 dimensions: *anxiety, depression, anger, excitement,* and *happiness.* Participants are asked to respond to 22 emotions on a Likert-type 5-point scale ranging from 0 (*not at all*) to 4 (*extremely*). The questionnaire was successfully used to assess emotions athletes recalled in the context of sports (Vast et al., 2010). Face, factorial, and construct validities have been examined on hundreds of athletes. A CFA indicated that the 22 items and 5-factorial structure resulted in a satisfactory model-data fit, $RCFI = 0.93$, $RMSEA = 0.07$, and very strong factorial loadings. Concurrent validity was tested through the SEQ correlations with the BRUMS (Terry et al., 1999, 2003a) and TOPS (Thomas et al., 1999), resulting in low (0.20-0.30) to moderate (0.30-0.50) correlations. The SEQ shared 77% items with the PNA, 50% with the POMS (McNair et al., 1971), 27% with the BRUMS, and 23% with the PANAS. Cronbach's $\alpha$ internal consistency reliability coefficients were anxiety ($\alpha = 0.87$), depression ($\alpha = 0.82$), anger ($\alpha = 0.84$), excitement ($\alpha = 0.81$), and happiness ($\alpha = 0.88$). The EC in this study measured participants' dispositional tendency for emotional contagion (a stable trait), while the SEQ measured state emotional responses at four time points: at baseline (pre-game) and following each of the three difficulty level stages.

**Collective Efficacy** (CE; Bandura, 1997). Collective Efficacy is measured by one item according to Bandura (1997) and Feltz and Chase (1998) postulations. Participants are asked to rate their perceptions about their team capacity in performing most effectively on the task. Specifically, they are asked: "How confident are you in your team's ability to serve orders on time?" on a scale ranging from 0 (*not sure at all*) to 100 (*very sure*). The team's collective efficacy score consists of the mean ratings of its team members.

**Communication Strategies: Content and Frequency**. Communication *content* was categorized into six statement types: uncertainty (UNC), action (ACT), acknowledgment (ACK), factual (FAC), non-task (NTS), and emotional/motivational (EMO) (Lausic et al., 2009).

*Frequency* was determined as the proportion of talk time relative to the total play time, expressed as a percentage. For instance, if the accumulated talk time of the team members in level 1 was 60 seconds and the level duration was 120 seconds, the frequency of talk time during that level was calculated as follows: (60 seconds / 120 seconds) × 100%. Therefore, in this example, the frequency of talk time during level number 1 was 50%.

Communication analysis included all team members' utterances, including the trained confederate. While the confederate followed prescribed communication patterns appropriate to each experimental condition, they retained flexibility to respond naturally to emerging gameplay situations. This approach preserved ecological validity by allowing authentic team interactions while ensuring systematic differences in supportive versus neutral communication styles between conditions. The confederate's communications thus served dual purposes: establishing the intended experimental climate and contributing to the team's overall communication patterns as would occur in naturalistic team settings.

**Team Task Performance**. Game performance was assessed by using the game performance scores and star ratings provided by Overcooked! 2. The game platform calculates performance scores as the number of successfully completed orders, the speed at which orders are fulfilled, and any penalties incurred for mistakes or unserved orders. In this study, all teams played three difficulty levels of Overcooked! 2 with increasing task complexity. Difficulty levels were operationally defined by: (a) number of simultaneous tasks required, (b) complexity of recipes, (c) presence of environmental obstacles, and (d) time pressure relative to order complexity. The selected levels were:

1. Level '1-1' (Easy): Features simple recipes with static kitchen layout. Requires basic coordination among team members and lasts 150 seconds.

2. Level '1-6' (Moderate): Introduces moving platforms and multi-step recipes. Presents more complex challenges requiring stronger communication and task allocation strategies and lasts 250 seconds.

3. Level '3-2' (Hard): Combines complex recipes with dynamic environmental hazards. Demands high levels of coordination, communication, and adaptability from team members to successfully complete the task and lasts 240 seconds.

**Intervention**

The study employed a trained confederate to ensure consistency in motivational climate communication styles across teams and to minimize potential biases. Critically, the confederate delivered identical task-related communications in both conditions (e.g., "we need to wash plates," "the onion is burning," "pass me the tomato"), with the key manipulation being the addition of supportive feedback in the PS condition versus the absence of such feedback in the NU condition. This approach ensured that the only systematic difference between conditions was the presence or absence of positive motivational climate elements, while keeping all task-relevant information constant. The confederate was an experienced player and shared a deep understanding of the study's objectives. The confederate received extensive training on the specific communication style to be exhibited in the PS and NU experimental conditions.

The training process involved several stages. First, the confederate was provided with detailed scripts outlining the key phrases, tone, and verbal cues associated with each experimental condition. He/she also participated in multiple practice sessions, during which he/she role-played the assigned communication behaviors with the research team. These sessions were reviewed to provide feedback and ensure consistency in the confederate's performance.

The confederate was instructed to maintain a consistent level of gameplay proficiency across all teams to ensure that any observed differences in team performance are primarily attributable to the motivational climate communication style rather than the confederate's gaming skills.

**Positive-Supportive (PS) Condition** - In the positive-supportive (PS) experimental condition, the confederate was engaged in verbal communication style that fostered a supportive and encouraging team environment. The confederate offered praise and acknowledgment for teammates' efforts and successes, using phrases such as "great job!", "nice move!", and "we're doing well, keep it up!" to reinforce positive performance. Even when faced with challenges or setbacks, the confederate maintained an optimistic tone, expressing confidence in the team's ability to overcome obstacles with statements like, "we can do this, just need to keep focusing on our goals". The confederate also actively acknowledged and appreciated teammates' contributions, using phrases like "thanks for covering that task," and "I appreciate your help with...". To promote a sense of unity and collaboration, the confederate employed inclusive language, such as "let's work together on this," and "we've got each other's backs."

**Neutral-Unsupported (NU) Condition** - In the neutral experimental condition, the confederate maintained a balanced and task-focused communication style. The confederate avoided both overtly positive and negative statements, instead focusing on objective, factual communication related to the game tasks. The confederate used neutral phrases such as "the order is ready", "we need more ingredients", or "time is running out". When addressing teammates' actions, the confederate used matter-of-fact statements like "that's done" or "that's not completed yet." The confederate neither praised nor criticized team performance, instead offering neutral observations such as "we finished that level" or "we didn't meet the goal this time." The confederate's tone remained even and unemotional throughout the game, neither encouraging nor discouraging teammates.

**Procedure**

This study was approved by the Institutional Review Board of *Masked* University, School of Psychology (ethical clearance number: P_2024097, approval date: 30.4.2024). All procedures were performed in compliance with relevant laws and institutional guidelines. The study was conducted in a controlled laboratory setting, equipped with PlayStation 4 gaming console, a 65-inch screen projecting the gameplay, and video recording equipment. Upon arrival, players were greeted by the experimenter and asked to provide informed consent. The players were randomly assigned to teams of three and introduced to their teammates, including the confederate. The confederate participated as a regular team member throughout the study, completing all questionnaires alongside the other participants to maintain the experimental integrity. However, only data from the actual participants were included in the statistical analyses of questionnaire responses; the confederate's questionnaire data were discarded.

Players were given a brief 2-minute tutorial to familiarize themselves with basic game controls and mechanics. This duration was sufficient for learning fundamental controls (movement, picking up/dropping items, basic cooking actions) without providing extensive practice that might mask learning effects during the experimental trials.

The main experiment consisted of three levels of gameplay. The difficulty level of the game increased with each level, starting with an easy level (level '1-1') and progressing to moderate (level '1-6') and hard level (level '3-2'). During each level, the confederate engaged in his/her assigned motivational climate, consistently delivering either positive-supportive or neutral-unsupported messages to their teammates. The content of these messages was designed to be relevant to the gameplay and the team's performance, while adhering to the specific motivational climate assigned to the confederate. Teams were pre-assigned to ensure the gender balance and match (see "participants"). The gender of the confederate was always matching the gender of the players in each team. After completing each of the three difficulty-level tasks, players have administered the SEQ and CE to assess changes in team emotion profile and collective efficacy respectively.

Throughout the gameplay sessions, the team communications were recorded using audio and video technologies. The recordings were analyzed using the Eldadi and Tenenbaum (2025) protocol to assess communication strategies and behaviors. Upon completion of the three levels' tasks, players were debriefed about the true purpose of the study and the role of the confederate. They were given the opportunity to ask questions and provide feedback about their experience. Finally, players were thanked for their participation.

**Data Analysis**

Data analyses aimed at testing the effects of motivational climate on team communication, performance, emotional states, and collective efficacy across increasing task difficulty levels. Communication data were analyzed using descriptive statistics and chi-square tests to examine content differences between motivational climates across difficulty levels. Effect sizes were calculated using Cohen's d coefficients and 95% confidence intervals (CI).

For quantitative variables, mixed repeated-measures multiple analyses of variance (mixed RM MANOVAs) were used to examine the effects of motivational climate (between-subjects) and task difficulty (within-subjects) on performance metrics, followed by univariate ANOVAs for specific outcomes. Emotional states were analyzed using mixed-design

MANOVA with motivational climate and emotional dimensions as factors. Collective efficacy was analyzed using mixed repeated-measures ANOVA.

Assumptions were verified using Box's test for covariance matrices and tests of normality. Levene's tests were used to examine variance homogeneity. Cohen's d effect sizes and 95% confidence intervals were calculated for significant effects ($p < .05$). When confidence intervals crossed zero, interpretations acknowledged the uncertainty of effects.

## Results

### Dispositional Emotional Contagion (EC) and Age Comparisons

To ensure that the players in the two experimental conditions were similar in age and dispositional emotional contagion, two independent t-test were performed. A non-significant age effect was noted between the PS ($M = 30.60$, $SD = 5.71$) and NU conditions ($M = 31.70$, $SD = 6.23$), $t(38) = -0.58$, $p = .56$. The second independent sample t-test revealed non-significant effect of the experimental condition on dispositional emotional contagion, $t(38) = 0.18$, $p = .86$ ($M = 42.20$, $SD = 5.00$ for PS condition vs. $M = 41.95$, $SD = 3.73$ for the NU condition).

Analysis of the covariance matrices across motivational climates using Box's test revealed no significant differences ($p > .05$) for all variables in the study. Furthermore, examination of the data distribution indicated no violations of normality assumptions, with acceptable levels of skewness and kurtosis observed throughout.

### Communications Analysis

A comprehensive analysis of player communication across 20 gameplay sessions yielded 5,824 recorded statements (23,335 words total), with 12,686 words (54.4%) expressed under the positive and supportive (PS) condition and 10,649 words (45.6%) under the neutral and unsupported (NU) condition.

#### *Motivational Climate Effects on Communication Content*

RM ANOVA GLM analysis using multi-variate approach examining the total number of statements by task difficulty level ($k = 3$), content categories ($k = 6$) and experimental conditions ($k = 2$), revealed a significant main effect for content category, $Wilks' \lambda = .12$, $F(5, 14) = 21.01$, $p < .001$, $\eta p^2 = .88$. Across the three difficulty levels and two experimental conditions, the predominant communication type was factual statements ($M = 30.20$ statements, $k = 1,823$), followed by action statements ($M = 27.45$, $k = 1,647$), uncertainty statements ($M = 16.75$, $k = 1,005$), emotional/motivational statements ($M = 11.57$, $k = 694$), acknowledgment statements ($M = 8.52$, $k = 511$), and non-task related statements ($M = 2.40$, $k = 144$).

The interaction between motivational climate and communication content category effect also reached significance, $Wilks' \lambda = .45$, $F(5, 14) = 3.48$, $p = .03$, $\eta p^2 = .55$, indicating that the contents of communication differed significantly in the PS and NU climates. None-parametric $\chi^2$ analysis and $\varphi$ coefficients performed within each difficulty level confirmed the differences between the two climates in communication content. Specifically, the analyses revealed a significant $\chi^2$ values observed at the easiest difficulty level, $\chi^2 = 42.24$, $df = 5$, $p < .001$, $\varphi = .192$, $n = 1147$, moderate level, $\chi^2 = 53.43$, $df = 5$, $p < .001$, $\varphi = .149$, $n = 2407$, and the hardest level, $\chi^2 = 48.36$, $df = 5$, $p < .001$, $\varphi = .146$, $n = 2270$.

In the PS climate, a significantly higher proportion of statements were factual ($M = 34.03$, $SD = 5.89$ vs. $M = 26.37$, $SD = 6.21$, $d = 1.27$, 95% CI [0.67, 1.86] – large effect) and action-oriented ($M = 30.30$, $SD = 5.45$ vs. $M = 24.60$, $SD = 5.78$, $d = 1.01$, 95% CI [0.47, 1.56] – large effect) compared to the NU climate. The PS climate also resulted in a higher amount of acknowledgment statements ($M = 10.00$, $SD = 3.12$ vs. $M = 7.03$, $SD = 3.45$, $d = 0.90$, 95% CI [0.38, 1.43] – large effect) and notably more emotional/motivational statements ($M = 15.50$, $SD = 4.20$ vs. $M = 7.63$, $SD = 3.89$, $d = 1.94$, 95% CI [1.19, 2.70] – very large effect). Conversely, the NU climate was characterized by significantly more uncertainty statements (*M* = 18.67, *SD* = 4.90 vs. *M* = 14.83, *SD* = 4.32, d = -0.83, 95% CI [-1.34, -0.32] – medium to large effect) and non-task-related statements (*M* = 3.27, *SD* = 1.45 vs. *M* = 1.53, *SD* = 1.05, *d* = -1.39, 95% CI [-2.03, -0.75]). Figure 2 presents the means and SDs of the number of statements made for the six content categories by task-difficulty level and motivational climate (positive-supportive vs. neutral-unsupported).

**Figure 2**

*Mean and SDs of statements per round (SpR) made for each communication content category and experimental condition (positive/supported vs. neutral/unsupported)*

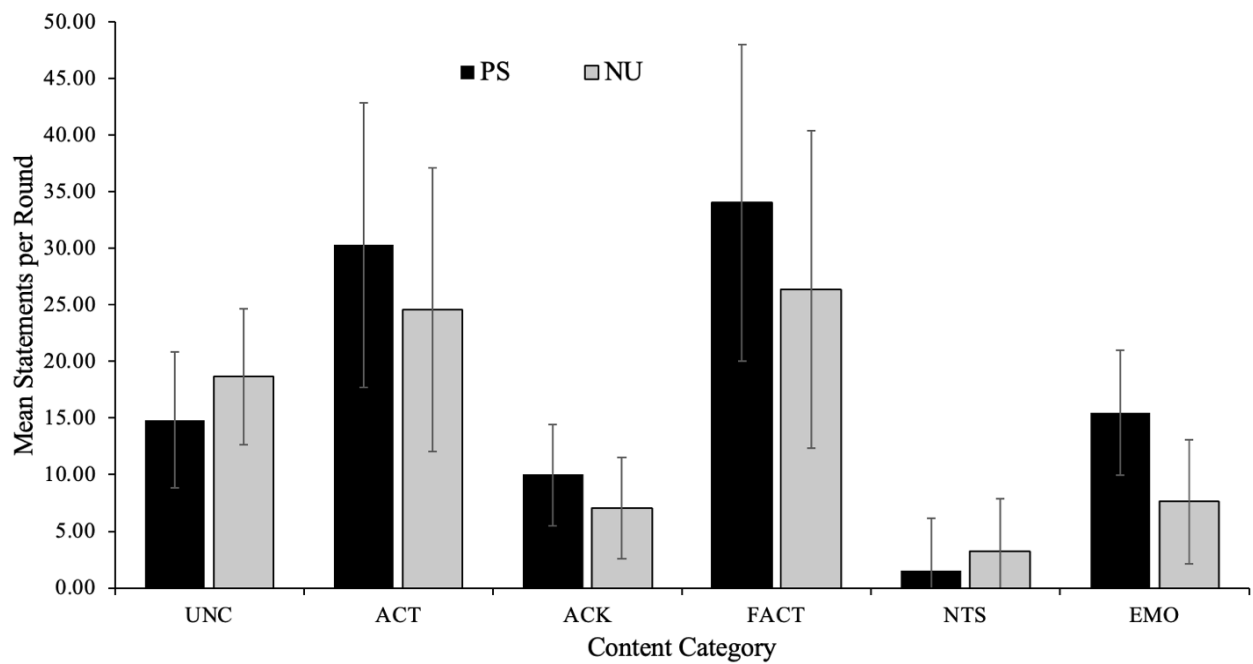

Note. UNC = Uncertainty, ACT = Action, ACK = Acknowledgment, FACT = Factual, NTS = Non-task, EMO = Emotional/motivational.

Performing the test of within-subject effects, the three-way interaction among the motivational climates, content categories, and task-difficulty level, was significant, *F*(10, 180) = 4.35, *p* = .02. As task difficulty increased, participants in both climates shifted their communication content, though these shifts were more pronounced in players in the PS climate.

At the easy task difficulty level, players in the PS climate maintained higher proportions of factual communication statements (*M* = 17.10, *SD* = 10.25) than players in the NU climate (*M* = 16.50, *SD* = 10.69, *d* = 0.06, 95% CI [-0.82, 0.93]) – negligible effect. Players in the PS condition used more action-oriented statements (*M* = 17.50, *SD* = 11.46) compared to the NU condition (*M* = 16.10, *SD* = 8.65, *d* = 0.14, 95% CI [-0.74, 1.02]) – negligible effect. The most notable difference appeared in emotional/motivational statements, where PS climate players expressed substantially more content (*M* = 13.70, *SD* = 8.31) than NU climate players (*M* = 6.80, *SD* = 5.07, *d* = 0.99, 95% CI [0.07, 1.91]) – large effect. Conversely, players in the NU climate displayed significantly higher levels of uncertainty statements (*M* = 10.90, *SD* = 4.01) compared to participants in the PS climate (*M* = 5.40, *SD* = 2.63, *d* = -1.62, 95% CI [-2.58, -0.65]) – large effect. For acknowledgment statements, PS climate players (*M* = 5.50, *SD* = 3.14) and NU climate players (*M* = 6.00, *SD* = 3.71, *d* = -0.14, 95% CI [-1.02, 0.73]) showed similar usage patterns. Non-task related communication was minimal in both conditions, with slightly higher occurrence in the NU climate (*M* = 2.60, *SD* = 4.97) than in the PS climate (*M* = 2.00, *SD* = 2.11, *d* = -0.16, 95% CI [-1.03, 0.72]) – negligible effect.

At the moderate task difficulty level, players in the PS climate maintained higher proportions of factual communication statements (*M* = 44.20, *SD* = 17.89) than players in the NU climate (*M* = 33.30, *SD* = 17.16, *d* = 0.62, 95% CI [-0.28, 1.51]) – moderate effect. The difference in action-oriented communication became more pronounced, with PS climate players using considerably more such statements (*M* = 38.80, *SD* = 15.47) than NU climate players (*M* = 27.50, *SD* = 15.81, *d* = 0.72, 95% CI [-0.18, 1.62]) – moderate to large effect. Similarly, the PS climate resulted in a greater use of emotional/motivational statements (*M* =

16.00, *SD* = 7.29) than in the NU climate (*M* = 7.50, *SD* = 4.97, *d* = 1.36, 95% CI [0.39, 2.33]) – large effect. PS climate players also showed increased use of acknowledgment statements (*M* = 12.50, *SD* = 7.35) compared to NU climate players (*M* = 8.10, *SD* = 7.46, *d* = 0.59, 95% CI [-0.30, 1.49]) – moderate effect. Conversely, players in the NU climate displayed higher levels of uncertainty statements (*M* = 22.80, *SD* = 8.61) compared to participants in the PS climate (*M* = 20.40, *SD* = 8.91, *d* = -0.27, 95% CI [-1.15, 0.60]) – small effect, and more non-task related communication (*M* = 4.30, *SD* = 9.79) than PS climate players (*M* = 1.40, *SD* = 1.96, *d* = -0.42, 95% CI [-1.30, 0.47]) – moderate effect. These shifting patterns reflect different adaptive strategies employed by participants as they confronted increasingly challenging tasks.

  At the highest task difficulty level, players in the PS climate maintained higher proportions of factual communication statements (*M* = 40.80, *SD* = 18.19) than players in the NU climate (*M* = 29.30, *SD* = 15.51, *d* = 0.68, 95% CI [-0.22, 1.58] – moderate effect. The difference in action-oriented communication remained substantial, with PS climate players using more such statements (*M* = 34.60, *SD* = 15.06) than NU climate players (*M* = 30.20, *SD* = 13.49, *d* = 0.31, 95% CI [-0.57, 1.19]) – small to moderate effect. Similarly, the PS climate resulted in a greater use of emotional/motivational statements at the highest difficulty level (*M* = 16.80, *SD* = 6.99) than in the NU climate (*M* = 8.60, *SD* = 6.42, *d* = 1.22, 95% CI [0.26, 2.18]) – large effect. PS climate players maintained higher use of acknowledgment statements (*M* = 12.00, *SD* = 4.55) compared to NU climate players (*M* = 7.00, *SD* = 3.83, *d* = 1.17, 95% CI [0.22, 2.12]) – large effect. Conversely, players in the NU climate displayed increased uncertainty in their communication as task difficulty increased, with higher levels of uncertainty statements (*M* = 22.30, *SD* = 10.11) compared to participants in the PS climate (*M* = 18.70, *SD* = 7.36, *d* = -0.41, 95% CI [-1.29, 0.47]) – moderate effect. NU climate players also demonstrated slightly higher non-task related communication (*M* = 2.90, *SD* =

4.80) than PS climate players (*M* = 1.20, *SD* = 1.14, *d* = -0.49, 95% CI [-1.38, 0.40]) – moderate effect. These shifting patterns reflect different adaptive strategies employed by participants as they confronted increasingly challenging tasks. The full descriptive statistics (*M* and *SD*) for the significant interaction effect among content category, task difficulty-level, and motivational climate is presented in Figure 3a,b,c.

**Figure 3**

*Mean and SDs of number of statements per round (SpR) for each communication content category, experimental condition (positive/supported vs. neutral/unsupported) and task difficulty-level*

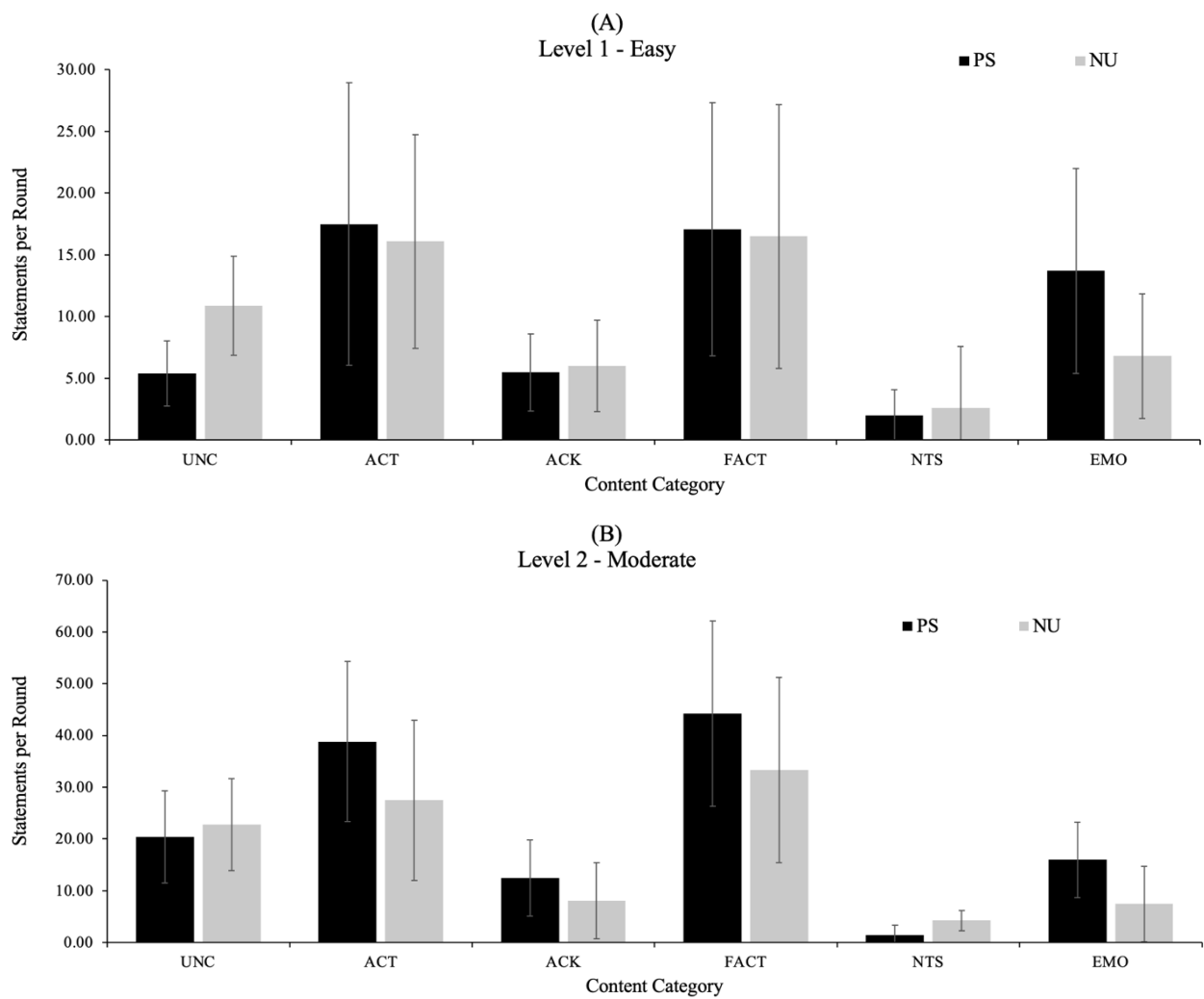

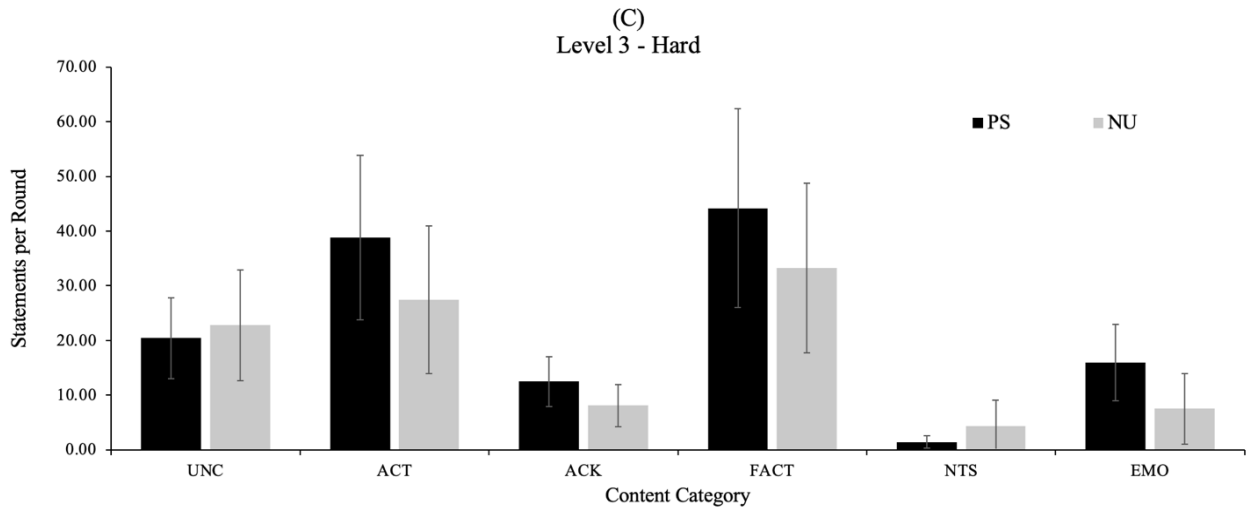

Note. UNC = Uncertainty, ACT = Action, ACK = Acknowledgment, FACT = Factual, NTS = Non-task, EMO = Emotional/motivational.

*Talk Time Percentage*

Analysis of talk time percentage revealed a significant main effect of task-difficulty level, *Wilks' λ* = .59, $F(2, 17)$ = 5.96, $p$ = .011, $\eta p^2$ = .41. Across all 20 gameplay sessions, the proportion of play time spent communicating increased progressively with task difficulty: level 1 (easy) = 43.90%, level 2 (moderate) = 51.16%, and level 3 (hard) = 52.53%. The interaction between motivational climate and task difficulty level for talk time percentage was non-significant, *Wilks' λ* = .82, $F(2, 17)$ = 1.84, $p$ = .19, $\eta p^2$ = .18, indicating that players in the two motivational climates used similar increases in talk time as difficulty increased. Figure 4 presents the means and SDs of talk time percentage by task difficulty and motivational climates.

**Figure 4**

*Means and SDs of talk time percentage by task difficulty level and experimental condition*

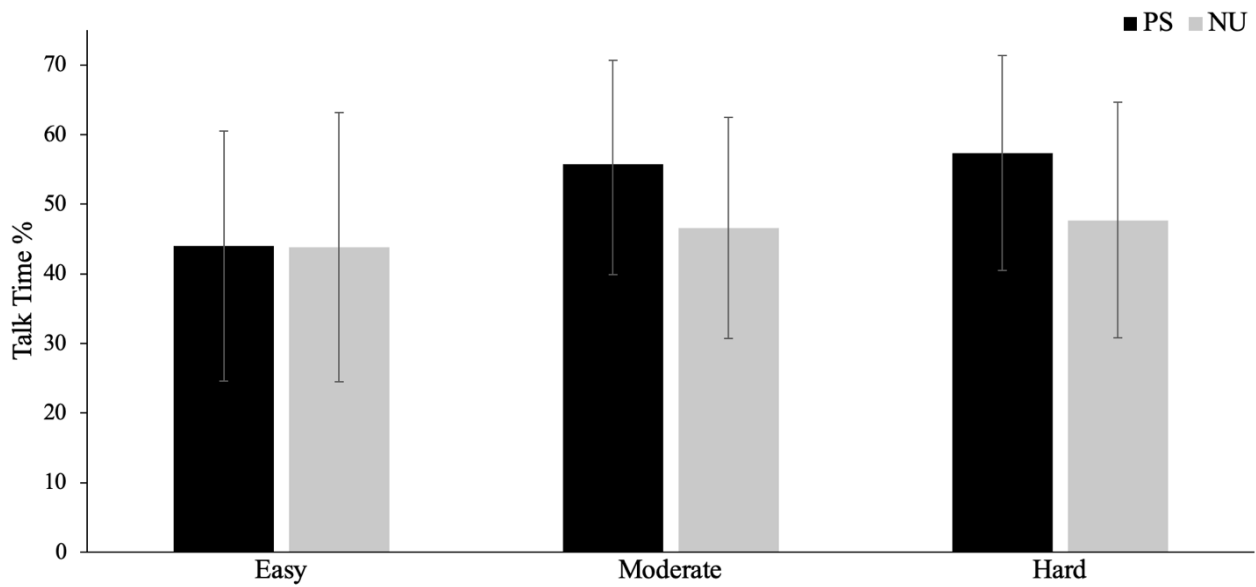

**Motivational Climate Effects on Teams' Performance**

Performance was measured using three key metrics, namely, *score*, *meal orders completed,* and *meal orders failed* across three escalating difficulty levels and two motivational climates (PS vs NU). Analyses were conducted using RM ANOVs GLM with task difficulty level as a WS repeated measures, and motivational climate as a BS factor.

*Performance Score*

A mixed RM GLM analysis revealed a significant main effect of task difficulty level on performance score, *Wilks' λ* = .02, $F(2, 17)$ = 348.38, $p < .001$, $\eta p^2$ = .98. Players in both motivational climates declined in performing the task as task difficulty level increased. Moreover, a significant interaction between motivational climate and task difficulty level emerged, *Wilks' λ* = .53, $F(2, 17)$ = 7.42, $p = .005$, $\eta p^2$ = .47. At the easiest difficulty level, players in the PS condition achieved significantly higher scores ($M$ = 911.20, $SD$ = 193.35) than players in the NU condition ($M$ = 744.70, $SD$ = 121.99), representing a large effect size

($d$ = 1.04, 95% *CI* [0.12, 1.96]). At the moderate difficulty level, the PS players continued to outperform the NU players (*M* = 420.70, *SD* = 234.54 vs. *M* = 353.10, *SD* = 171.39), maintaining a moderate difference ($d$ = 0.33, 95% *CI* [-0.55, 1.21]). By the hardest level, while players in the PS condition scored nominally lower than the NU condition (*M* = 222.10, *SD* = 143.09 vs. *M* = 231.70, *SD* = 126.88), the difference was minimal and not statistically significant ($d$ = -0.07, 95% *CI* [-0.95, 0.81]).

### Meal Orders Completed

A significant main effect of task difficulty level on orders completed was evident, *Wilks' λ* = .01, $F(2, 17)$ = 578.35, $p < .001$, $\eta p^2$ = .99. Teams completed significantly fewer orders as task difficulty level increased, with mean completed orders declining from easy (*M* = 25.10, *SD* = 4.64) to moderate (*M* = 8.35, *SD* = 2.72), and further decreasing at hard level task (*M* = 4.55, *SD* = 1.88).

A significant difficulty level by motivational climate interaction emerged for orders completed, *Wilks' λ* = .50, $F(2, 17)$ = 8.35, $p$ = .003, $\eta p^2$ = .50. Players in the PS condition maintained a higher completion rate at the easy task difficulty level, with this advantage diminishing as task difficulty level increased. Specifically, at the easy level, the PS players significantly outperformed the NU players (*M* = 27.40, *SD* = 4.67 vs. *M* = 22.80, *SD* = 3.46, $d$ = 1.11, 95% CI [0.18, 2.04] - large effect). At the moderate level, players in the PS condition completed slightly more meal orders than their NU counterparts (*M* = 8.70, *SD* = 3.34 vs. *M* = 8.00, *SD* = 2.06, $d$ = 0.25, 95% CI [-0.63, 1.13] - small effect). By the final hardest level, the performance gap disappeared, with meal order completion becoming nearly identical in the two conditions (*M* = 4.40, *SD* = 2.22 vs. *M* = 4.70, *SD* = 1.57, $d$ = -0.15, 95% CI [-1.02, 0.72] - negligible effect).

### Meal Orders Failed

Given that failed orders at the easiest difficulty level failed to affect the final score,

the analysis pertained only to moderate and hard difficulty levels. The analysis revealed that task difficulty level significantly influenced the number of failed orders, *Wilks' λ* = .54, *F*(1, 18) = 15.58, *p* < .001, $\eta p^2$ = .46, indicating that as game difficulty increased, team players in both motivational climates failed to complete more orders. Specifically, the average number of failed orders more than doubled from the moderate level (*M* = 1.00, *SD* = 1.08) to the hard level (*M* = 2.35, *SD* = 1.39), regardless of communication condition. Moreover, the analysis revealed a non-significant effect of motivational climate on the number of failed meal orders, *Wilks' λ* = .95, *F*(1, 18) = 1.05, *p* = .32, $\eta p^2$ = .05. Players in both motivational climates increased in failed orders as task difficulty increased.

**Motivational Climate Effects on Emotional States**

The analysis revealed a significant interaction effect between motivational climate and emotional dimension, *Wilks' λ* = .73, *F*(12,27) = 3.30, *p* < .05, $\eta p^2$ = .27, indicating that players' emotional profiles varied significantly between the two motivational climates. Figure 5 illustrates the differences in emotional dimensions between the two motivational climates.

**Figure 5**

*Mean scores for five emotional dimensions (Anxiety, Dejection, Excitement, Anger, and Happiness) in Positive-Supportive (PS) and in Neutral-Unsupported (NU) experimental conditions.*

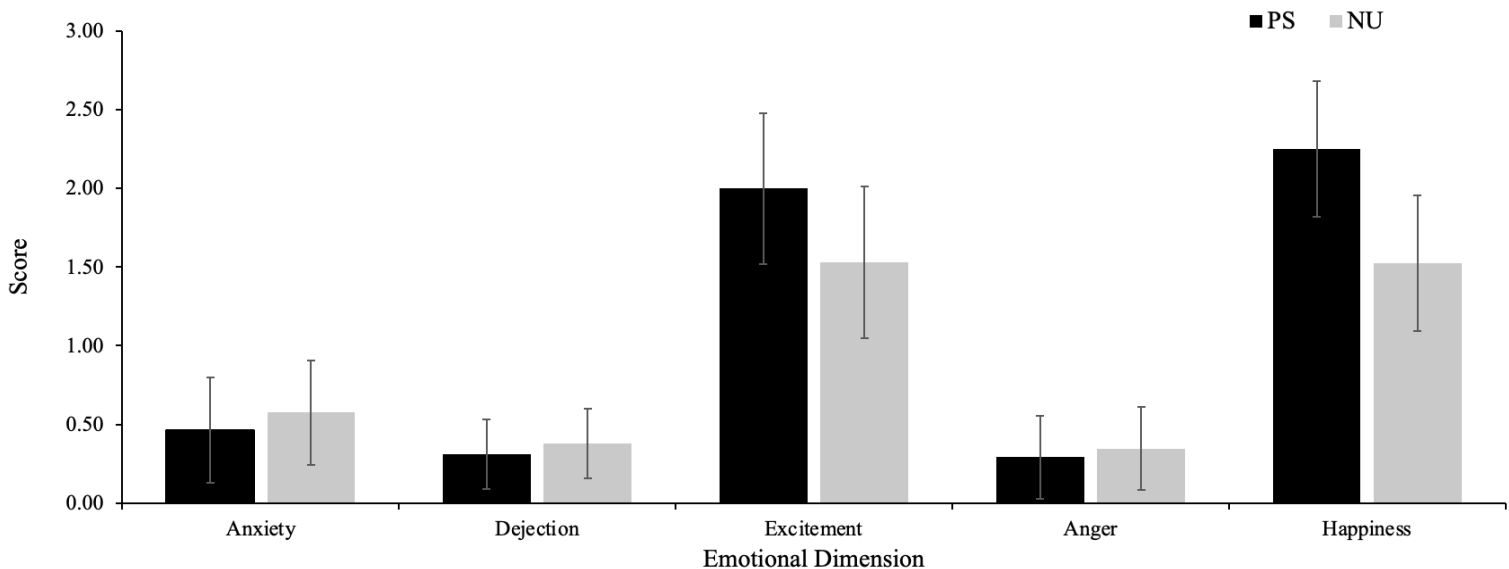

Across the three levels of task difficulty, *Anxiety* was reported to be higher in players within the NU climate ($M = 0.57$, $SD = 0.33$) than in players within the PS climate ($M = 0.46$, $SD = 0.33$), representing a small effect, $d = 0.33$, 95% *CI* [-0.91, 1.58]. Similarly, *Dejection* was reported to be higher in players competing in the NU climate ($M = 0.38$, $SD = 0.22$) compared to players competing in the PS climate ($M = 0.31$, SD = 0.22), also indicating a small effect, $d = 0.32$, 95% *CI* [-0.93, 1.56]. *Excitement* was reported to be substantially higher in the PS climate condition ($M = 2.00$, $SD = 0.48$) than in the NU condition ($M = 1.53$, $SD = 0.48$), representing a large effect, $d = -0.97$, 95% *CI* [-2.29, 0.34]. *Anger* feeling were slightly higher in the NU climate players ($M = 0.35$, $SD = 0.26$) compared to the PS climate players ($M = 0.29$, $SD = 0.26$), $d = 0.21$, 95% *CI* [-1.03, 1.46]. The most pronounced difference was reported for *Happiness*, with significantly stronger happiness in the PS climate players (*M* = 2.25, *SD* = 0.43) compared to the NU climate players (*M* = 1.52, *SD* = 0.43), demonstrating a large effect, *d* = -1.69, 95% *CI* [-3.13, -0.24].

**Collective Efficacy (CE)**

The Analysis revealed a significant main effect of task difficulty, *Wilks' λ* = .56, *F*(3, 36) = 9.35, *p* < .001, $\eta p^2$ = .44, indicating that collective efficacy perceptions declined as task difficulty level increased. However, the interaction between motivational climate and task difficulty failed to be significant, *Wilks' λ* = .95, *F*(3, 36) = 0.68, *p* = .57, $\eta p^2$ = .05.

Descriptive statistics for each task difficulty level revealed a pattern of decreasing collective efficacy. Prior to the easy task (first measurement), players in the PS climate reported higher collective efficacy than their team members in the NU climate (*M* = 7.50, *SD* = 2.04 vs. *M* = 7.10, *SD* = 1.62, *d* = 0.22, 95% *CI* [-0.40, 0.84] - small effect). After completing the easy task difficulty level, team members in both climates reported their highest collective efficacy scores (*M* = 7.80, *SD* = 1.58 vs. *M* = 7.60, *SD* = 1.27, *d* = 0.14, 95% *CI* [-0.48, 0.76] - negligible effect). A notable decline in collective efficacy was reported following the moderate task difficulty level, with players in the PS climate maintaining higher efficacy beliefs than those in the NU climate (*M* = 6.55, *SD* = 2.14 vs. *M* = 5.60, *SD* = 2.04, *d* = 0.45, 95% *CI* [-0.18, 1.08] - moderate effect). After completing the hardest task difficulty level, the players in both climates reported a slight recovery in collective efficacy scores compared to the moderate task difficulty level, with the PS climate players again reporting higher scores than the players in the NU climate (*M* = 6.70, *SD* = 2.03 vs. *M* = 6.30, *SD* = 1.87, *d* = 0.20, 95% *CI* [-0.42, 0.82] - small effect).

## Discussion

This study examined how different communication climates (e.g., positive-supportive (PS) versus neutral-unsupported (NU)) affected team dynamics in the collaborative gaming environment of Overcooked! 2. We measured communication patterns, emotional states, collective efficacy, and performance across increasing task difficulty levels. Our findings demonstrate that motivational climate, operationalized through supportive versus neutral communication patterns, was associated with differences in multiple team processes and outcomes, though these effects varied with task difficulty.

**Motivational Climate Effects Across Multiple Team Dimensions**

Our results revealed that motivational climate influenced multiple aspects of team functioning, including communication patterns, emotional states, collective efficacy, and performance. The analyses examined the effect of motivational climate on each outcome separately. The PS climate was associated with more adaptive patterns across most measured variables compared to the NU climate, suggesting that supportive communication environments may create conditions that benefit multiple team processes simultaneously.

PS teams used higher proportions of action-oriented and factual statements that previous research has associated with effective coordination (Eldadi & Tenenbaum, 2025; Freeman & Wohn, 2017; Lausic et al., 2009). While these variables showed parallel differences between conditions, our analyses cannot determine whether communication patterns mediated the relationship between motivational climate and other outcomes. The co-occurrence of adaptive communication, positive emotions, and higher efficacy in PS teams suggests these factors may reinforce each other, consistent with theoretical frameworks proposing interdependence among team processes (Lausic et al., 2009; Freeman & Wohn, 2017).

Simultaneously, team members in the PS climate maintained significantly higher levels of excitement and happiness while experiencing lower levels of anxiety, dejection, and anger than players in the NU climate. This "emotional advantage" aligns with Barsade's (2002) emotional contagion theory, where positive emotions spread throughout the team, creating a psychological environment that may support positive team interactions.

Moreover, team members in the PS climate maintained consistently higher collective efficacy beliefs throughout the three task difficulty levels, which co-occurred with positive emotional states and effective communication patterns. The fact that these variables were both higher in PS teams aligns with Bandura's (1997) efficacy theory, where emotional state and efficacy beliefs are theoretically linked and together shape performance outcomes.

The findings of this study also revealed that task difficulty level moderated these effects in somewhat unexpected manner. Contrary to our hypothesis, PS teams significantly outperformed NU teams at the easy task difficulty level, but this advantage diminished as difficulty increased, eventually disappearing at the hardest difficulty level. This performance outcome pattern reveals that the relationship between psychological processes and performance outcomes is not linear but is moderated by task complexity. At lower difficulty levels, the psychological and communicative advantages conferred by PS climate were associated with superior performance outcomes. However, as technical demands intensified at the highest difficulty level, these psychological advantages were counterbalanced by the complexity of the task.

Importantly, even as objective performance differences disappeared at the hardest task difficulty level, players operating in the PS climate maintained their advantage in communication patterns, emotional profile, and collective efficacy. The dissociation between psychological processes and performance outcomes at high difficulty levels provides important insights into the boundary conditions of motivational climate effects. As Klimoski and Mohammed (1994) noted, when tasks become extremely complex, the advantages from team shared mental models may reach their limits, and technical skill factors become more determinative of performance outcomes.

**Implications for Team Functioning in Dynamic Environments**

Our findings have implications for capturing team functioning in dynamic environments. The consistent pattern of differences between PS and NU teams across multiple variables, e.g., communication, emotions, and efficacy, suggests that motivational climate may have broad effects on team dynamics. While we cannot determine how these variables influence each other, their co-occurrence indicates that fostering supportive communication environments might benefit multiple aspects of team functioning simultaneously.

This integrated perspective helps reconcile seemingly contradictory findings in previous research. Some studies have emphasized the primacy of communication in determining team outcomes (Eldadi & Tenenbaum, 2025; Lausic et al., 2009; Marlow et al., 2018; Musick et al., 2021), while others have highlighted emotional processes (Eldadi et al., 2023) or efficacy beliefs (Goddard et al., 2004) as critical determinants. Our findings suggest that these perspectives are complementary rather than competing, as we observed differences in multiple components simultaneously. The consistent pattern of results across communication, emotional, and efficacy variables indicates these factors may be related, though our design cannot establish how they might influence each other.

The dissociation between psychological processes and performance outcomes at high difficulty levels also has important theoretical implications. The fact that players in the PS climate maintained their advantage in communication patterns, emotional profile, and collective efficacy even as performance differences disappeared suggests that psychological processes may operate on a different timescale than performance outcomes, as proposed by Mesmer-Magnus et al. (2021). While immediate performance may be constrained by task difficulty, the psychological advantages from positive motivational climates may provide long-term benefits for team adaptability and resilience that would manifest over extended collaboration periods.

**Limitations and Future Directions**

Several limitations must be considered when interpreting the results of this study. First, the laboratory setting with the video game Overcooked! 2, while offering excellent experimental control, may not fully replicate the complexity and sustained pressure of real-world team contexts. Future research could extend these findings to naturalistic team environments such as sports teams, emergency response units, or workplace project teams, following approaches suggested by Eccles and Tenenbaum (2004) for studying team coordination in field settings.

Second, the use of a confederate to manipulate motivational climate, while ensuring experimental consistency, may not capture the more organic development of motivational climates in real teams. Future studies could examine how motivational climates emerge naturally from peer-to-peer interactions rather than being established by a single leader figure.

Third, the relatively short duration of the team interaction in this study limits our understanding of how motivational climate effects might evolve over extended periods of collaboration. Longitudinal designs would provide valuable insights into the sustainability of positive motivational climate effects and potential adaptation processes over time.

Our communication findings reflect team-level patterns that include the confederate's contributions. While this approach captures how different communication climates manifest in actual team interactions, it means that observed differences between conditions partially reflect the experimental manipulation itself. The proportion of supportive versus neutral statements necessarily differs between conditions by design. However, examining how participants' communications adapted in response to these different climate conditions provides insight into the contagion effects of communication styles within teams. More research might separately analyze confederate and participant communications to disentangle stimulus from response patterns.

While our analyses examined relationships between motivational climate and various team outcomes, we did not test a sequential mediation directly. We recommend employing appropriate statistical techniques such as sequential mediation analysis to verify the simultaneous influences from climate through communication and emotions to performance.

In summary, our findings reveal that motivational climate was associated with differences in communication patterns, emotional states, efficacy beliefs, and performance outcomes, with these effects moderated by task difficulty. While these variables showed consistent patterns of difference between conditions, more research using sequential statistical methods is required to uncover their mutual interrelationships.